\documentclass[12pt]{article}
\usepackage{amsmath}
\usepackage{physics}
\usepackage{graphicx,psfrag,epsf}
\usepackage{enumerate}
\usepackage[round]{natbib}
\usepackage{url} 
\usepackage{float}
\usepackage{caption}
\usepackage{subcaption}

\usepackage{glossaries}
\usepackage{multirow}

\newacronym{sej}{SEJ}{Structured Expert Judgementy}

\newacronym{ara}{ARA}{Adversarial Risk Analysis}

\newacronym{eum}{EUM}{Expected Utility Maximisation}

\newacronym{aru}{ARU}{Axiomatic Random Utility}

\newacronym{mnl}{MNL}{Multinomial logit}

\newacronym{dm}{DM}{Decision Maker}

\newcommand{\blind}{0}

\DeclareMathOperator*{\argmax}{argmax} 
\newcommand{\SP}{{\cal P}}

\usepackage{tikz}
\usetikzlibrary{arrows,automata,backgrounds,decorations,fit,patterns,petri,positioning,shadows,shapes}

\pgfdeclarepatternformonly{stripes}{\pgfpoint{0cm}{0cm}}{\pgfpoint{1cm}{1cm}}{\pgfpoint{1cm}{1cm}}{
  \foreach \i in {0.125, 0.375,...,0.875}{
    \pgfpathmoveto{\pgfpoint{\i cm}{0cm}}
    \pgfpathlineto{\pgfpoint{1cm}{1cm - \i cm}}
    \pgfpathlineto{\pgfpoint{1cm}{1cm - \i cm + 0.125cm}}
    \pgfpathlineto{\pgfpoint{\i cm - 0.125cm}{0cm}}
    \pgfpathclose
    \pgfusepath{fill}
    \pgfpathmoveto{\pgfpoint{0cm}{\i cm}}
    \pgfpathlineto{\pgfpoint{1cm - \i cm}{1cm}}
    \pgfpathlineto{\pgfpoint{1cm - \i cm - 0.125cm}{1cm}}
    \pgfpathlineto{\pgfpoint{0cm}{\i cm + 0.125cm}}
    \pgfpathclose
    \pgfusepath{fill}
  }
}

\begin{document}

\def\spacingset#1{\renewcommand{\baselinestretch}%
{#1}\small\normalsize} \spacingset{1}


\if0\blind
{
  \title{\bf Forecasting Adversarial Actions Using Judgment Decomposition-Recomposition}
  \author{Yolanda Gomez, DevStat, Spain\\ 
    Jesus Rios, IBM Research, USA\\ 
    David Rios Insua, ICMAT-CSIC, Spain\\    
    Jose Vila, Universidad de Valencia, Spain }
    \date{ } 
  \maketitle

} \fi

\if1\blind
{
  \bigskip
  \bigskip
  \bigskip
  \begin{center}
    {\LARGE\bf Title}
\end{center}
  \medskip
} \fi

\bigskip

\begin{abstract}

In domains such as homeland security, cybersecurity and competitive marketing, it is frequently the case that 
analysts need to forecast adversarial actions that impact the problem of interest.
Standard structured expert judgement 
elicitation techniques may fall short as they do not explicitly take into account intentionality. We present a decomposition technique based on adversarial risk analysis followed by
a behavioral recomposition using discrete choice models that 
facilitate such elicitation process and illustrate its performance 
through behavioral experiments.

\end{abstract}

\vfill

\newpage
\spacingset{1.45} 


\section{Introduction}
\label{sec:cap3introduction}

\acrfull{sej} elicitation \citep{cooke1991experts} is a major ingredient within decision analysis \citep{clemen2013making}. A significant feature in the practice of this discipline, already 
  acknowledged in  \citet{raiffa1968decision}'s seminal book, is its emphasis in decomposing complex problems into smaller pieces that are easier to understand and then recomposing the piecewise solutions to tackle the global problem.

Indeed, when applying decision analysis, 
its methodology suggests solving complex decision making problems through maximum expected utility (MEU). In doing so, one avoids direct comparison of alternatives, which may be cognitively intricate and prone to bias, specially in presence of uncertainty and multiple objectives. For this,
the problem is structured by identifying alternatives, uncertainties and objectives;  the decision maker's beliefs and preferences are 
assessed;  and, then, the  MEU alternative is found. 
The value of such type of analysis is assessed in e.g.\ \citet{watson1978valuation}.
				
Preference elicitation assessment also uses decomposition and
recomposition, given that it is usually difficult to compare consequences of alternatives, 
specially in presence of multiple conflicting attributes. A typical approach is to search for a decomposable functional form for the utility function \citep[often additive, multiplicative or multilinear, e.g.][]{gonzalez2018utility} and then assess the component utilities and weights to later recompose the global utility function whose expected value must be maximised. \citet{ravinder1991random} and \citet{ravinder1992random} provide theory showing the advantages of undertaking such decompositions. 

Finally, belief elicitation benefits as well from decomposition through the argument of \emph{extending the conversation} \citep{Ravinder1988}.
\citet{tetlock2015superforecasting} call it \emph{Fermitisation} and consider it a key element
for the success of their {\em super-forecasters}. Rather than directly assessing the probability of an outcome, one finds a conditioning partition and assesses the probabilities of the outcome given the conditioning events. From these, and the probabilities of such events, the law of total probability recomposes the desired unconditional probability of the outcome \citep{lawrence2006judgmental}. \citet{andradottir1997choosing,andradottir1998analysis} provide a methodological framework to validate this approach, empirically evaluated in e.g.\ \cite{macgregor2001decomposition}. Based on this data, \cite{armstrong2006findings} identifies the forecasting situations under which judgmental decomposition showed improvement in accuracy.

Broadly speaking, decomposition (and the corresponding
recomposition) uncovers the complexity underlying direct assessment, eliminating the burden on experts to perform sophisticated modelling in their heads. This simplifies complex cognitive tasks 
and mitigates reliance on heuristics that introduce bias, promoting that experts actually analyse the relevant problem \citep{montibeller2015biases}. Decompositions typically entail more assessments, albeit simpler and more meaningful, leading to improved judgements and decisions at large. 

Here we focus on developing and validating experimentally a  decomposition-recomposition 
strategy to support \acrshort{sej} when forecasting adversarial actions. 
There are two main uses for this. 
For the first one, in line with \citet{kadane1982subjective} and \citet{raiffa1982art}, 
assume we employ decision analysis to support a decision maker in dealing with game theoretic problems; this naturally leads to try to forecast adversarial actions.
Concerning the second use, in many settings in contexts such as security, counterterrorism, intelligence, cybersecurity, or 
competitive marketing, experts will face problems in which they need to deal with probabilities referring to actions to be potentially carried out by opponents. As an example, an important percentage of the questions posed to experts in \citet{chen2016validating}  referred to  adversarial actions (e.g.\ {\em Will Raja Pervez Ashraf resign or otherwise vacate the office of Prime Minister of Pakistan before 1 April 2013?} or \emph{Will the Palestinian group Islamic Jihad significantly violate its cease-fire with Israel before 30 September 2012?}).

We could think of using standard \acrshort{sej} tools, as in \citet{dias2019elicitation} or  \citet{hanea2021expert}, to deal with such problems. However,
\citet{keeney2007modeling} cogently argues that knowledge about the adversaries' beliefs and preferences may not be that precise as it would require them to reveal their judgements,
which is doubtful in security and other domains where information
is concealed. Alternatively, we study here whether
 \acrfull{ara} decompositions \citep{banks2015adversarial} serve better for such adversarial forecasting purposes. \citet{dias2020adversarial} argue theoretically that \acrshort{ara} may be used 
advantageously as a decomposition-recomposition
strategy for adversarial forecasting. Here, we assess it from a behavioural 
perspective, using 
a behavioral experiment \citep{vilacybeco}. For this, we integrate \acrshort{ara} with discrete choice models  \citep{thurstone1927law,marschak1959binary,mcfadden1973conditional,train} to provide a behavioral perspective on such methods, adopting an asymmetric prescriptive-descriptive approach 
in the sense of \cite{raiffa1982art}:
we use a {\em prescriptive} view for the supported 
agent assessing and decomposing her uncertain view of  her adversary's
preferences and beliefs and later recompose them with the aid of {\em descriptive} discrete choice models to obtain improved adversarial forecasts. In such a way, we provide a novel behavioral perspective 
for adversarial 
belief assessment.

After sketching the \acrshort{ara} approach to decomposition (Section \ref{sec:ara}), and briefly recalling the theoretical arguments to justify it, together with its integration with discrete choice models for recomposition, Section \ref{sec:cap3rationale} presents the experiments undertaken.  Section \ref{sec:exp_results} analyses results drawing conclusions.  Section \ref{sec:cap3discussions} ends up with a discussion.


\section{Decomposition-recomposition for adversarial 
forecasting}
\label{sec:ara}

Let us start by outlining the role of \acrshort{ara} as a \acrshort{sej} decomposition method 
and, then, describe its integration with discrete choice models for recomposition purposes. 


\subsection{Predicting adversarial actions}
\label{sec:predicting}

\begin{figure}[htbp!]
    \centering
    \begin{subfigure}[b]{0.39\textwidth}
    \centering
        \begin{tikzpicture}[->, >= stealth', shorten >= 1pt, auto, semithick, scale = 0.85,
                      transform shape,
                      roundnode/.style={circle, draw, minimum width = 1cm},
                      squarednode/.style={rectangle, draw, minimum width = 1cm, minimum height = 1cm}]
                      
            \node[roundnode   ](A0)  [                                              ]{};             
            \node[            ](DA01)[above right = 0.5cm and 2cm of A0]{};
        	\node[            ](DA02)[below right = 0.5cm and 2cm of A0]{}; 
        	\node[            ](empty)[below = 1.5cm of A0]{};

            \path(A0)  edge[out =  45, in = 180] node [align=center, 
                       font=\footnotesize\linespread{0.8}\selectfont]{not act\\$1-p_A$   }(DA01)
        		       edge[out = -45, in = 180] node [align=center, below, pos=0.3, yshift=-5pt, font=\footnotesize\linespread{0.8}\selectfont]{act\\$p_A$}(DA02);
          \end{tikzpicture}
        \caption{Direct assessment}
    \end{subfigure}
    \hfill
    \begin{subfigure}[b]{0.6\textwidth}
    \centering
        \begin{tikzpicture}[->, >= stealth', shorten >= 1pt, auto, semithick, scale = 0.85,
                      transform shape,
                      roundnode/.style={circle, draw, minimum width = 1cm},
                      squarednode/.style={rectangle, draw, minimum width = 1cm, minimum height = 1cm}]
                      
            \node[squarednode ](A)   [                                      ]{};
        	\node[            ](DA1) [above right = 0.5cm and 2cm of A,   label = right:${u=p_B}$]{STATUS QUO};
        	\node[roundnode   ](DA2) [below right = 0.5cm and 2cm of A,                            ]{};
        	\node[            ](W)   [above right = 0.5cm and 2cm of DA2, label = right:${u = 1}$]{BEST};
        	\node[            ](L)   [below right = 0.5cm and 2cm of DA2, label = right:${u = 0}$]{WORST};
        
            \path(A)   edge[out =  45, in = 180] node [align=center, 
                       font=\footnotesize\linespread{0.8}\selectfont]{not act}(DA1)
        		       edge[out = -45, in = 180] node [align=center, below, pos=0.3, yshift=-5pt, font=\footnotesize\linespread{0.8}\selectfont]{act}(DA2)
        		 (DA2) edge[out =  30, in = 180] node [align=center, yshift=-8pt,
        		       font=\footnotesize\linespread{0.8}\selectfont]{succeed\\$p_C$}(W)
        		       edge[out = -30, in = 180] node [align=center, below, pos=0.3, yshift=-5pt, font=\footnotesize\linespread{0.8}\selectfont]{not\\succeed\\ $1-p_C$}(L);
         \end{tikzpicture}
        \caption{Assessment through ARA decomposition}
    \end{subfigure}
  \caption{Two views on adversarial forecasts.} 
  \label{fig:adec}
\end{figure}
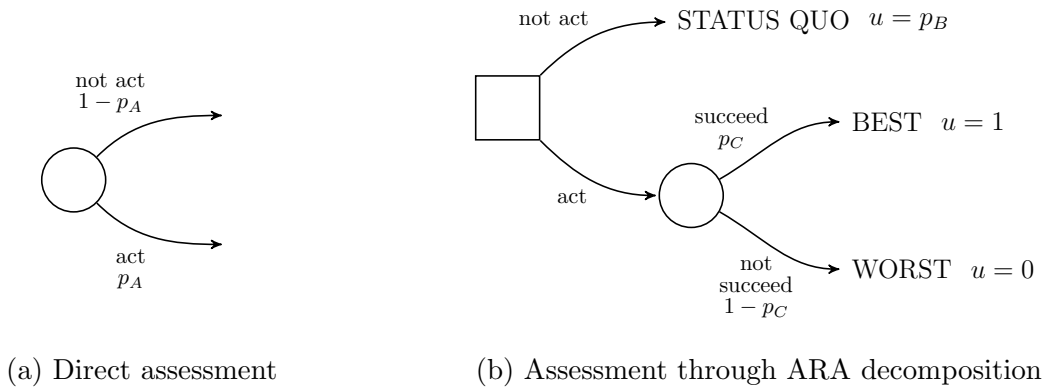

Our focus is on the second type of uses proposed for 
forecasting adversarial actions,
 as Figure \ref{fig:adec} reflects. This serves us also to sketch the experiments 
 presented in Section \ref{sec:cap3rationale}. The left panel represents the uncertainty of interest for an agent $D$ (she), who aims at forecasting whether an adversary (he, $A$) will act, which will happen with probability $p_A$. This 
 corresponds to its direct assessment, as in  \cite{lewis}. The right panel reflects the decision problem faced by the adversary: if he does not act, the status quo remains and he attains utility $u$, whose value 
will be designated $p_B$; if he acts and succeeds, which happens with probability $p_C$, he gets the best result, with associated utility $u=1$; however, if he fails, he would get the worst result with associated utility $u=0$. Recall, e.g. \cite{gonzalez2018utility}, that
utility functions are affine unique and thus we may fix two of its values with no loss of generality.

\begin{figure}[htbp!]
    \centering
    \begin{subfigure}[b]{0.24\textwidth}
        \includegraphics[width=\textwidth]{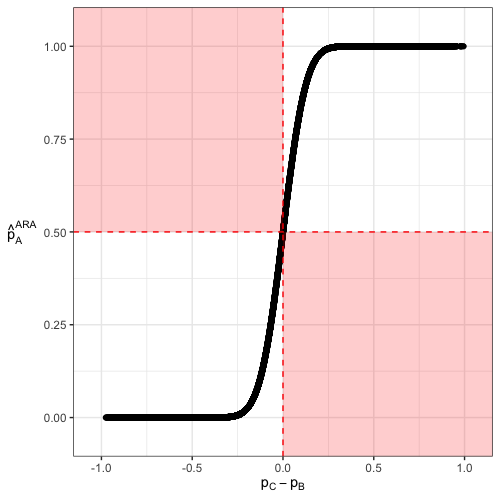}
        \caption{$\sigma^2 = 0.1$}
        \label{fig:ara_sd_01}
    \end{subfigure}
    \begin{subfigure}[b]{0.24\textwidth}
        \includegraphics[width=\textwidth]{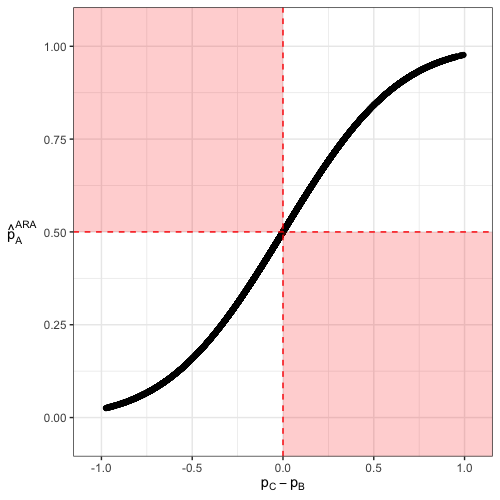}
        \caption{$\sigma^2 = 0.5$}
        \label{fig:ara_sd_05}
    \end{subfigure}
    \begin{subfigure}[b]{0.24\textwidth}
        \includegraphics[width=\textwidth]{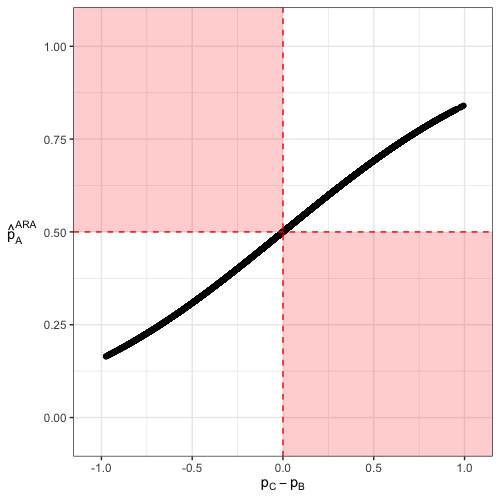}
        \caption{$\sigma^2 = 1$}
        \label{fig:ara_sd_1}
    \end{subfigure}
    \begin{subfigure}[b]{0.24\textwidth}
        \includegraphics[width=\textwidth]{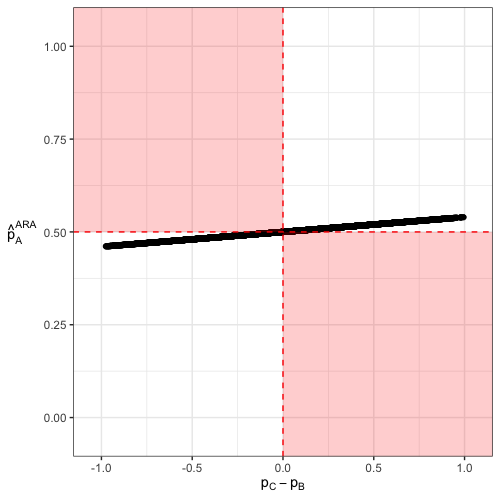}
        \caption{$\sigma^2 = 10$}
        \label{fig:ara_sd_10}
    \end{subfigure}
    \caption{ARA recompositions $\hat{p}^{ARA}_A$ for different values of the uncertainty parameter $\sigma^2$.}
    \label{fig:ARA_recompositions}
\end{figure}

Assume for the moment that the adversary maximises EU in his 
decision making. Should we know $u=p_B$ and $p_C$, as the expected utility of 
the available 
actions {\em not act} and {\em act} are, respectively, 
$p_B$ and $p_C$, we would predict that the adversary 
will {\em act} (i.e., $p_A=1$) if $p_B <p_C$ and will {\em not act}
($p_A=0$), otherwise. However, 
in line with \cite{keeney2007modeling}'s comments,
as the adversary is not available for 
elicitation purposes, there is uncertainty about such elements. 
If we model it through distributions $P_B$ (a random 
utility) and $P_C$ (a random probability), we would actually have uncertainty about the adversary's decision and 
estimate $\widehat{p}^{ARA}_A = Pr ( P_B < P_C)$. For example, if the uncertainty about $p_B$ and $p_C$ is modeled through normally distributed random probabilities $P_B$ and $P_C$ centered respectively around $p_B$ and $p_C$
, then $P_C-P_B \sim N(p_C-p_B,\sigma^2)$, 
and the \acrshort{ara} estimate in this case would be
\begin{equation}\label{pARA} \hat{p}^{ARA}_A = Pr (P_C-P_B > 0) = 1-\Phi\left(\frac{p_B-p_C}{\sigma^2}\right), \end{equation} 
where the parameter $\sigma^2$ can be interpreted as a measure of agent $D$'s uncertainty on her assessments of $p_B$ and $p_C$ and 
$\Phi$ designates the cdf of  the standard normal distribution. Figure~\ref{fig:ARA_recompositions} shows \acrshort{ara} estimates for different values of $\sigma^2$, as recompositions from point estimates $p_B$ and $p_C$. We see how the \acrshort{ara} recomposition is equivalent to \acrshort{eum} when $\sigma^2\longrightarrow 0$ and how $\hat{p}^{ARA}_A$ flattens around $0.5$ as the uncertainty about $p_B$ and $p_C$,    assessed through $\sigma^2$, increases.

Yet we  could argue that, again, as the adversary $A$ is not 
actually available, agent $D$  can consider that
 $A$ may apply other rules, beyond expected utility maximisation, to select an action \citep{schoemaker1982expected}. In general, $A$'s 
decision rule can be modeled as $\widehat{p}_A = F(p_B , p_C)$, where $F$ is a generic functional. Additional examples of 
specific forms for $F$ are 
next presented.



\subsection{Recomposing adversarial beliefs with 
discrete choice models}
\label{sec:recomposing}

 The introduction of $F$   
  allows us to consider alternative behavioural choice models to represent $D$'s beliefs about the decision rules used by her adversary. 
  These alternative decision rules could 
  emerge, for instance, from potential cognitive biases or heuristics affecting 
  the decision procedure of $A$, from $D$'s perspective.
  Such behavioural insights would be expected to improve the accuracy of the recomposed estimations of the adversary's probabilities of adopting his actions.  
 
 Inspired by Eq.~(\ref{pARA}), which relates to the probit
 discrete choice model \citep{train},
we additionally explore 
two alternatives stemming from such family 
of models. Their   
initial concepts were proposed by \citet{marschak1959binary} and \citet{block1959random}, and later developed in \citet{mcfadden1973conditional}. In these models, the decision  maker's utility is subject to random shocks. In a behavioural economics setting, such shocks 
are typically interpreted as a result of the agents' cognitive biases, of decision-making heuristics or of implementation errors in the decision process \citep{hess2018revisiting}. Utility maximization, together with empirically estimated shocks, leads to probabilistic
criteria for alternative choice.
  Note that although both \acrshort{ara} and discrete 
  choice modelling translate into probabilistic rules for adversary's action selection, the underlying uncertainty is different in both approaches: in \acrshort{ara} models, uncertainty about the actions to be selected is a consequence of the agents beliefs on their adversaries' probabilities and utilities; on top of it, in random utility models, the utility of each agent is intrinsically random. 

The probability distribution on the action to be chosen by 
the adversary in discrete choice  models can be obtained 
through constructive or axiomatic approaches. Initially,
 \citet{marschak1959binary} and \citet{mcfadden1973conditional}
 obtained the distribution by construction:
  the model is specified by defining an
a priori utility 
and  randomness is introduced by  making behaviour depend 
as a random function of this fixed utility \citep{busemeyer2014psychological}. As an example,
 MacFadden obtains a functional form for the choice probability, known as 
 multinomial logit model \citep{berbeglia}, also called exponential softmax function in the 
 machine learning literature \citep{bishop}, given by

\begin{equation}\label{ryan1}
    \widehat{p}^{MNL}_A= \frac{e^{u(i)}}{\sum_{j \in S}e^{u(j)}}, i \in S,
\end{equation}
where $u(i)$ is the agent's utility for alternative $i$ and 
$S$ is the discrete set of alternatives available.
The axiomatic treatment, initiated by Luce (1959), 
introduced a choice axiom (LCA) postulating 
how the probability of selecting an alternative 
is related to the probability of selecting such alternative from a larger set. The Axiomatic Random Utility (ARU) approach leads to a probability 
of selecting the $i$-th alternative from a set $S$ 
through  

\begin{equation}\label{ryan2}
    \widehat{p}^{ARU}_A = \frac{u(i)}{\sum_{j \in S}u(j)}, i \in S. 
\end{equation}

 Detailed discussions on the LCA validity may be seen  
 in \citet{luce1977choice} and \citet{pleskac2015decision}.
 \cite{guikema} use a version of this rule to forecast terrorist threats.

The behavioural recomposition models to be analysed empirically
 will integrate the ARU and MNL probabilistic choice rules $\widehat{p}^{ARU}_A$ and $\widehat{p}^{MNL}_A$ within \acrshort{ara} models. The accuracy of these behavioural recompositions will be compared and tested with that of standard \acrshort{ara} considering expected utility maximization. This comparison will provide valuable insights for developing effective \acrshort{sej} methodologies when forecasting adversarial actions.

\section{Validating behavioural ARA for SEJ decomposition-recomposition: rationale}
\label{sec:cap3rationale}

\subsection{Experiment design}
\label{sec:cap3design}

To validate the proposed \acrshort{sej} decomposition-recomposition method, a group of participants was recruited to elicit their beliefs on uncertain adversarial events within the controlled setting of a behavioural experiment,
as in e.g.\ \cite{vilafinance}, \cite{vila} or \cite{vilacybeco}.

A challenge for the  design was to build an user-friendly mechanism to help the participants reveal such beliefs in a reliable and consistent manner. To this end, the experiment focused on the analysis of 
uncertain events involving strategic decision-making in three domains: politics, consumer  products and sports. The specific questions employed 
were:
{\em Will Theresa May ask for elections in the next 30 days?};  {\em Will Coca Cola's CEO announce a new marihuana based drink in the next 30 days?}; {\em Will Rafael Nadal participate in a hard court competition in the next 30 days?}.\footnote{The experiment took place in April 2019, when 
the three questions were relevant globally.}
The design paid attention to the careful wording of the questions and the framing of the answering process used. 
The experiment was piloted with 10 subjects, who also participated in individual face-to-face debriefing interviews to confirm their understanding of the procedure and identify potential improvements.  

For each of the uncertain events, the subjects were required to complete four experimental tasks per forecasting question (12 tasks in total).  These 
 were presented sequentially 
 according to the following logic, recall Figure \ref{fig:adec}:  

\begin{itemize}
    \item {\em Task 1.} Beliefs about a well-known person, the adversary in our terminology, referred  to as \acrfull{dm}, launching  a strategic action in a given period of time, denoted $p_A$, as Figure \ref{fig:taska} shows.
        \item {\em Task 2.} Beliefs about the {\it minimum} success probability under which the \acrshort{dm} would launch the strategic action, from which we deduce $p_B$,
        based on the protocol shown in Figure \ref{fig:taskb}.
    Through this task, the subject is actually revealing his beliefs about the DM's utility of not acting, since utilities for acting and succeeding or failing are normalised to 1 (best outcome) and 0 (worst outcome), respectively.
           \item {\em Task 3}. Beliefs about the chances of the \acrshort{dm} succeeding if the strategic action is launched, denoted  $p_C$, as Figure \ref{fig:taskc}
           shows. 
    \item {\em Task 4}. Repetition of \textit{Task 1}. We ask again for the probability $p_D$ of the \acrshort{dm} launching the strategic action in the given period of time. This repetition allows for checking 
    whether a participant has changed her initial belief assessment $p_A$ after the exercise of reflecting about $p_B$ and $p_C$.
    \end{itemize}
As the three figures show, the participant 
actually assessed the required  probabilities in a scale from 0\%  to 100\%, in
    steps of 10\%, thus
 effectively identifying 
 intervals of width 10\%  where the participant believes that the corresponding probability lies. 
 
 Finally, we asked participants to self-assess their degree of knowledge in the thematic domain in which the \acrshort{dm} is considering to launch or not the strategic question as well as to report their socio-demographic profile. 


\subsection{Experiment implementation}
\label{sec:cap3experiment}

The face-to-face experiment was carried out in an experimental laboratory. It was run using a software specifically developed, thoroughly tested before the sessions to guarantee stability, usability and understandability. 
The experiment comprised four sessions, each with 24 participants, for a total of 96 subjects. At the beginning of each session, subjects were randomly located around semi-cubicles in a room. To mitigate presentation bias, the same facilitator led all sessions. He read the instructions and accompanied his speech with a slide projector to explain the type of questions
 that the subjects would have to answer through three examples, and the benefits that they would obtain as a function of the performance of their forecasts, since participation 
  entailed a variable payment depending on the probabilities assigned to the  events and their actual realisation 30 days after the session. They then undertook the actual tasks,
with a median duration of approximately half an hour, leaving about 10 minutes per topic.

\section{Validating behavioural ARA for SEJ decomposition-recomposition: results}
\label{sec:exp_results}

Our analysis covers the following stages.
 
\begin{enumerate}

    \item We first check for consistency in the respondents answers for tasks $1$ to $4$
    for the three questions described in Section~\ref{sec:cap3design}. 
    
    \item Next, we generate probabilistic forecast estimates through recompositions 
    based on $p_B$ and $p_C$, specifically based on expected utility maximization (EUM), axiomatic random utility (ARU), multinomial logit (MNL), and adversarial risk analysis (ARA). We also discuss the consistency of the derived probabilistic estimates.

    \item We then demonstrate whether the decomposed-recomposed probability assessments improve intuitive predictive capabilities 
    with the aid of proper scoring rules after the events took place. 
    
    \item Finally, we check whether the reflection 
    entailed by the decomposition-recomposition exercise
    leads to substantial changes concerning the probabilities assessed directly in Tasks 1 and 4. 
    
\end{enumerate}

At each stage, we use exploratory data analysis and, when necessary, support conclusions via (Bayesian) hypothesis testing~\citep{Giron}.
General conclusions are then drawn.

\subsection{Checking for consistency in respondents answers}
\label{sec:coherence}

Our first analysis is essentially exploratory and aims to reflect upon the 
global coherence of each participants' responses. 
Recalling the discussion in Sections \ref{sec:predicting} and \ref{sec:cap3design}, we can relate the assessment of $p_A$ (probability that the \acrshort{dm} chooses to {\em act}) with those of $p_C$ (what the \acrshort{dm} believes to be her success probability if she {\em acts}) and $u=p_B$ (elicited as the minimum success probability under which the \acrshort{dm} would {\em act}). Thus, we would expect that a respondent who knows with certitude that $p_B < p_C$ (a success probability higher than the minimum by which the \acrshort{dm} would {\em act}), will assess $p_A$ to be $1$. However, taking into account the actual uncertainty of the participant about the \acrshort{dm}'s judgements $p_B$ and $p_C$, we shall consider $p_A>1/2$ for the respondent assessments to be consistent. Moreover, the assessment of $p_A$ is done before the respondent is exposed to Tasks $2$ and $3$, where he goes through the exercise of decomposing his judgments. Thus, although we do not expect consistency to appear
necessarily in intuitive judgments such as $p_A$, we would when the respondent is given the opportunity to review his initial judgment after Tasks $2$ and $3$, and 
then provides a revised $p_D$ in Task $4$.

\begin{figure}[htbp!]
    \centering
    \begin{subfigure}[b]{0.45\textwidth}
        \includegraphics[width=\textwidth]{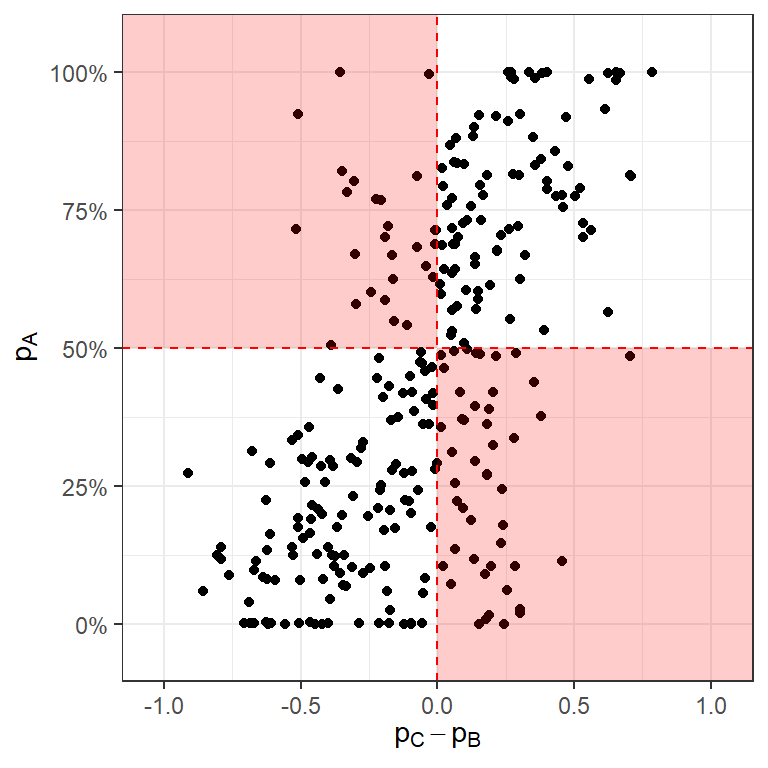}
    \end{subfigure}
    \caption{Exploring initial inconsistencies in the respondent's assessments.}
    \label{fig:coherence}
\end{figure}

Figure~\ref{fig:coherence} shows a view concerning the consistency of the respondent's initial assessments of $p_A$ in our experiment. The panel represents the scatter plot for the 288 (96 individuals, 3 questions) responses of $(p_C - p_B, p_A)$. Points in white areas, ($p_B < p_C$, $p_A>1/2$) and ($p_B < p_C$, $p_A<1/2$), represent consistent assessments in the sense described above, whereas those in red areas suggest inconsistent ones.
Observe the important number of observations in the red area. 
 
\pagebreak 

Table~\ref{tab:coherence} summarises the results, essentially conveying 70.2\% of consistent judgments (those on the diagonal) vs.\ 29.9\% of inconsistent ones (those off the 
diagonal) about $p_A$, providing strong empirical evidence supporting 
that the distributions of subjects considering that  $p_C<p_B$ and $p_C>p_B$ differ in terms of $p_A$ being greater or smaller than $1/2$.

\begin{table}[h!]
  \centering
    \begin{tabular}{ccccl} \hline
              & $p_C<p_B$ & $p_C>p_B$ &           \\ \hline
    $p_A<1/2$ & 41.0\%    & 17.7\%    & \\
    $p_A>1/2$ & 12.2\%    & 29.2\%    &            \\ \hline
    \end{tabular}
    \caption{Consistent and inconsistent initial judgments.}
    \label{tab:coherence}
\end{table}


Table~\ref{tab:agreementlevel} displays the
percentage of participants with 0 (full consistency), 1, 2 or 3 inconsistent judgments with regards to their initial assessments of $p_A$ on the three  questions 
in the  experiment.

\begin{table}[h!]
    \centering
    \begin{tabular}{cccc} \hline
           0         & 1           & 2           & 3                \\ \hline
           36.5\%  & 42.7\% & 15.6\%   &  5.2\%          \\ \hline
    \end{tabular}
    \caption{ 
    Percentage of participants depending on number of inconsistent judgments with regards to the initial assessment of $p_A$.}
    \label{tab:agreementlevel}
\end{table}

All in all, this suggests lack of consistency between the respondents' initial judgments about $p_A$ (their direct forecasts about whether the \acrshort{dm} will {\em act}) and the decomposed assessments 
recomposed afterwards to construct an alternative forecast. Nearly 64\% of the participants incurred at least once in decomposed assessments inconsistent with their initial intuitive judgments.   

A relevant question is whether respondents with inconsistent assessments in Task $1$ vs.\ Tasks $2$ and $3$ changed their initial $p_A$ judgment in Task $4$ to some revised 
$p_D$ judgement so as to resolve such inconsistency or, more
graphically, whether responses $(p_C - p_B, p_A)$ in the red areas in       Figure~\ref{fig:coherence} moved towards the appropriate white areas after being revised to $(p_C - p_B, p_D)$.
In the experiment, 30\% of respondents modified their judgments, and, of those,
12.8\% of the inconsistent judgments were made consistent after revision. 
On the other hand, 3.5\% of the initial consistent judgments were made inconsistent 
after the revision. Thus, we appreciate that the exposure of our subjects to elicitation Tasks $2$ and $3$ led to minor improvements in terms of resolving inconsistencies in their judgments.

\subsection{Inducing consistency through recomposition}
\label{sec:decomposition-recomposition}

As suggested in Section~\ref{sec:recomposing}, we can use discrete choice models to combine the elicited quantities $p_B$ and $p_C$ and produce an estimate $\hat{p}_A$ of the probability that the \acrshort{dm} will choose to {\em act}, following the 
rationale in Figure~\ref{fig:adec}. We shall consider the following recompositions

\begin{itemize}
    \item \acrfull{eum} recomposition
    \begin{equation*}
        \hat{p}^{EUM}_A = 
        \begin{cases}
            1   & \text{if } p_C > p_B\\
            1/2 & \text{if } p_C = p_B, \\
            0   & \text{if } p_C < p_B
        \end{cases}
    \end{equation*}
     
       \item \acrfull{mnl} recomposition, as in (\ref{ryan1})
    \begin{equation*}
        \hat{p}^{MNL}_A = \frac{e^{p_C}}{e^{p_C} + e^{p_B}}
    \end{equation*}
        \item \acrfull{aru} recomposition, as in (\ref{ryan2})
    \begin{equation*}
        \hat{p}^{ARU}_A = \frac{p_C}{p_C + p_B}.  
    \end{equation*}
        \item \acrfull{ara} recomposition, given by~(\ref{pARA}) 
        \begin{equation*} \hat{p}^{ARA}_A =  1-\Phi\left(\frac{p_B-p_C}{\sigma^2}\right). \end{equation*}         
        In our experiments, we assessed $p_B$ and $p_C$ when decomposing the judgment about $p_A$, but we did not directly ask for the variance $\sigma^2$ that measures the respondent's uncertainty on his/her assessments of $p_B$ and $p_C$. We indirectly estimated $\sigma^2$ by asking our subjects to self-assess their level of domain-knowledge with regards to each of the forecasting questions in a $5$-point Likert scale and defined a correspondence between $\sigma^2$ and the level of knowledge, with decreasing values of $\sigma^2$ as the level of knowledge increases. 
        
 \end{itemize}
 
\begin{figure}[htbp!]
    \centering
    \begin{subfigure}[b]{0.49\textwidth}
        \includegraphics[width=\textwidth]{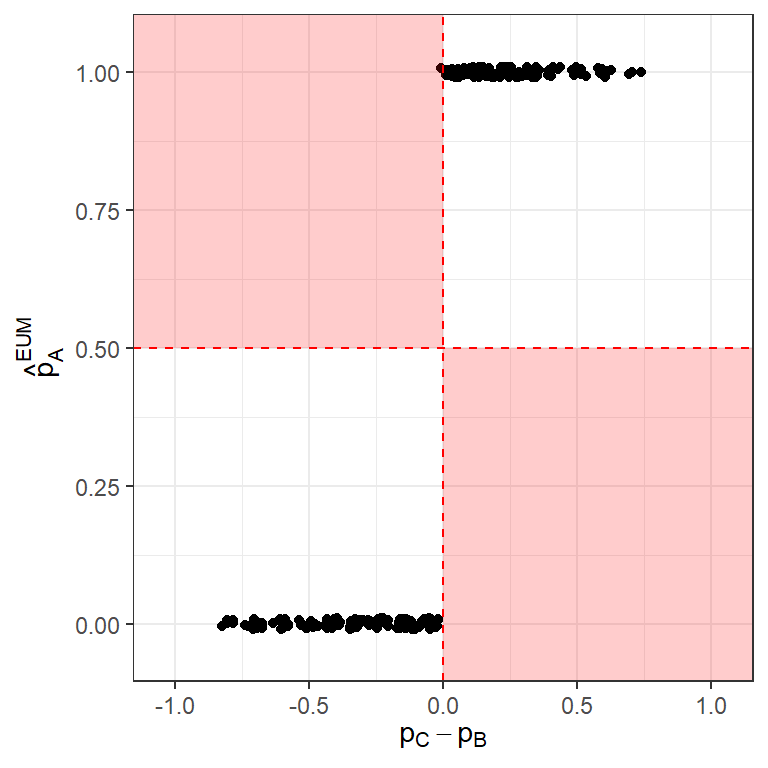}
        \caption{EUM}
        \label{fig:decompositioncoh_a}
    \end{subfigure}
    \begin{subfigure}[b]{0.49\textwidth}
        \includegraphics[width=\textwidth]{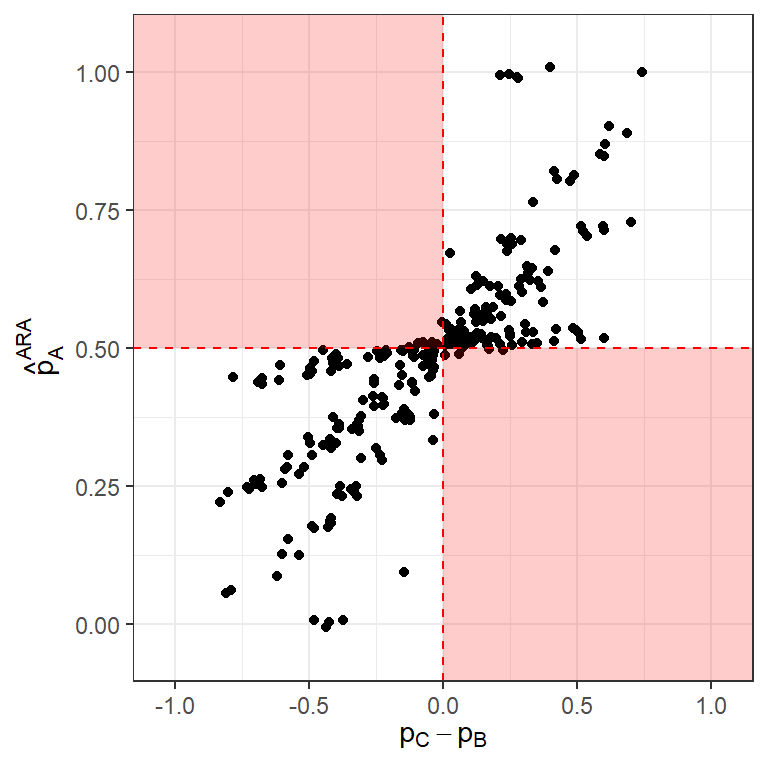}
        \caption{ARA}
        \label{fig:decompositioncoh_d}
    \end{subfigure}
    \begin{subfigure}[b]{0.49\textwidth}
        \includegraphics[width=\textwidth]{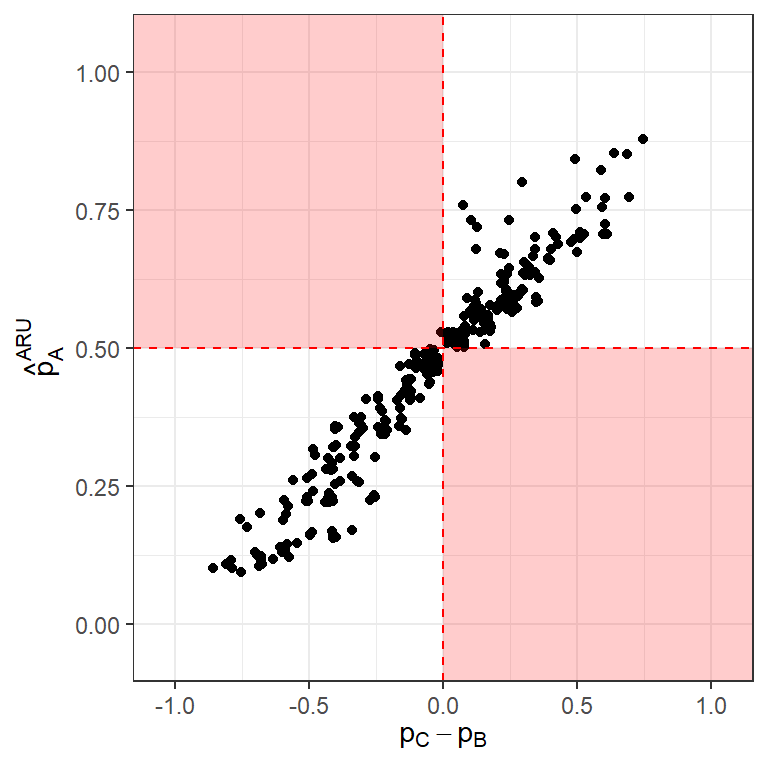}
        \caption{ARU}
        \label{fig:decompositioncoh_c}
    \end{subfigure}
    \begin{subfigure}[b]{0.49\textwidth}
        \includegraphics[width=\textwidth]{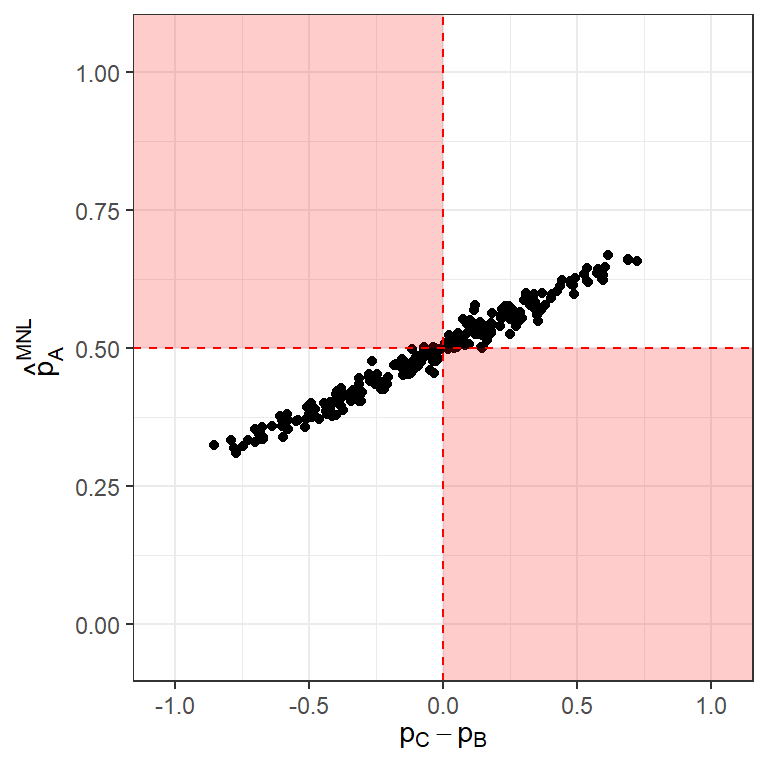}
        \caption{MNL}
        \label{fig:decompositioncoh_b}
    \end{subfigure}
    \caption{Recomposition of probabilities through EUM, ARA, ARU and MNL. Jittered to
    facilitate visualization.}
    \label{fig:decompositioncoh}
\end{figure}

Figure \ref{fig:decompositioncoh} represents the scatterplots of $p_C-p_B$ and the
 probability recompositions. It is easy to show theoretically that the \acrshort{eum}, \acrshort{mnl}, \acrshort{aru} and \acrshort{ara} recompositions induce probabilities $\hat{p}_A$ that are consistent with respect to their constituent assessments $p_B$ and $p_C$. Empirically, this is observed through the absence of ($p_C-p_B$, $\hat{p}_A$) 
 points in the red areas in Figure \ref{fig:decompositioncoh}. Thus, the use of any of these recompositions $\hat{p}_A$, instead of the direct judgments $p_A$, would ensure coherence with respect to the assessments obtained in the decomposition phase of our experiments.

\subsection{Predictive capabilities of the recompositions}
\label{sec:predictive_capabilities}

A comparison of the forecasts derived from the initial direct assessment of $p_A$,
   Figure~\ref{fig:coherence},  and our recompositions based on $p_B$ and $p_C$,  Figure~\ref{fig:decompositioncoh}, reveals the differences between $p_A$ and the (EUM, ARA, ARU, MNL) recompositions. However, this only tell us that recomposed probabilistic forecasts  are very different than direct ones, with the most extreme differences observed for the \acrshort{eum} recomposition. Hence, we next further analyse whether these  recompositions improve forecast
 accuracy. 

Thus, 
 we evaluate the recompositions introduced in Section~\ref{sec:decomposition-recomposition} in terms of their predictive performance 
 in our experiment and compare
them 
 with the direct forecasting judgments $p_A$ given by our respondents before exposing them to thinking in terms of decompositions in Tasks $2$ and $3$. These results, comparing direct intuitive forecasting judgments with those obtained using decomposition-recomposition techniques, will provide a justification for adopting this type of techniques.

To perform this evaluation we shall:

\begin{itemize}
    \item Take into account the actual imprecision in the assessments since, recall
     Section \ref{sec:cap3design}, the proposed tasks identified the required 
     probabilities within an interval of width 10\%. We acknowledge this 
     by modelling some uncertainty around such assessments through 
     distributions, see details in Appendix B.
    Such distributions will serve  for observation sampling whenever requested.
    \item The \cite{brier1950verification} score is used to assess the predictive capabilities of the models considered. Recall that for a single binary question 
    (that is with only two possible outcomes, an event occurs or not), this 
    score is $(f-o)^2$ where $f$ is the probability assigned to the event by a model and $o$ is 0 if the event did not happen and $1$, otherwise. The score range is 
     $[0,1]$. The smaller the score is, the better the predictive capability of the corresponding forecast. 
      For a non-informative respondent answering 0.5,
    the Brier score would be 0.25.
    
\end{itemize}

\begin{figure}
    \centering
        \begin{subfigure}[b]{0.8\textwidth}
            \includegraphics[width=\textwidth]{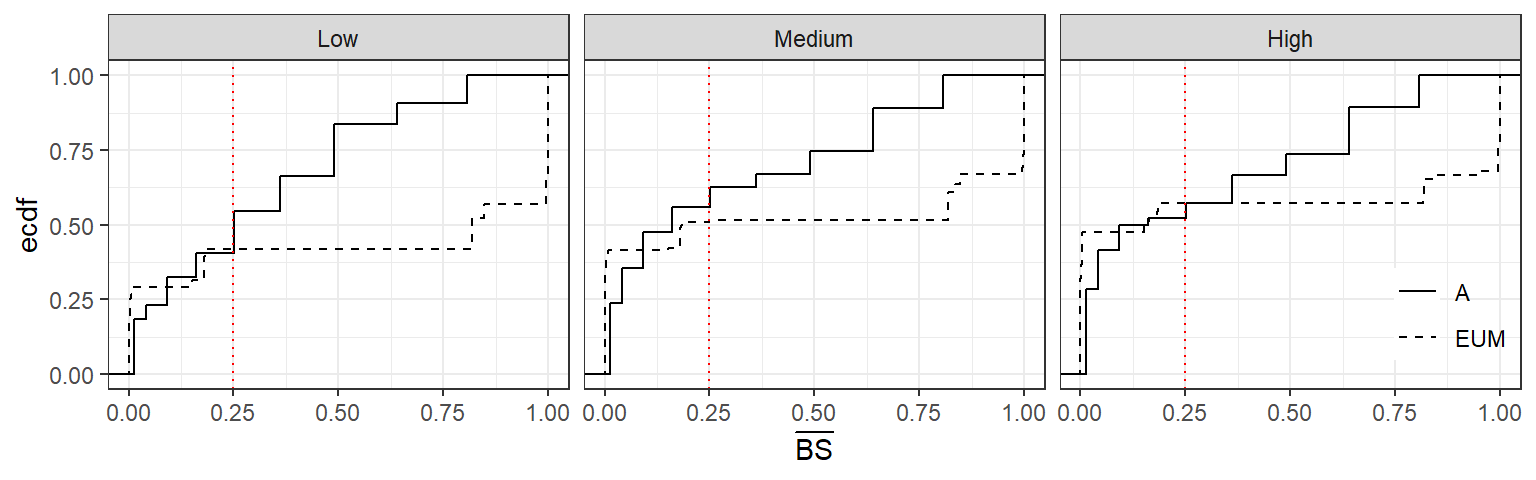}
            \caption{EUM}
            \label{fig:BS_comp_EUM}
        \end{subfigure} 
        \begin{subfigure}[b]{0.8\textwidth}
            \includegraphics[width=\textwidth]{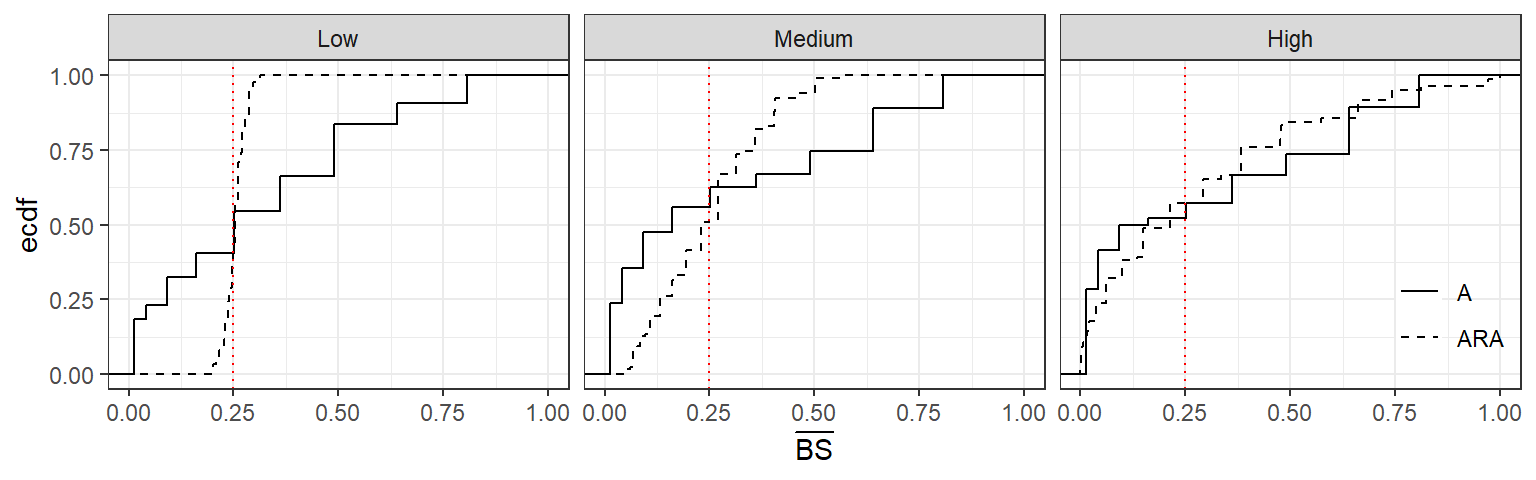}
            \caption{ARA}
            \label{fig:BS_comp_ARA}
        \end{subfigure} 
        \begin{subfigure}[b]{0.8\textwidth}
            \includegraphics[width=\textwidth]{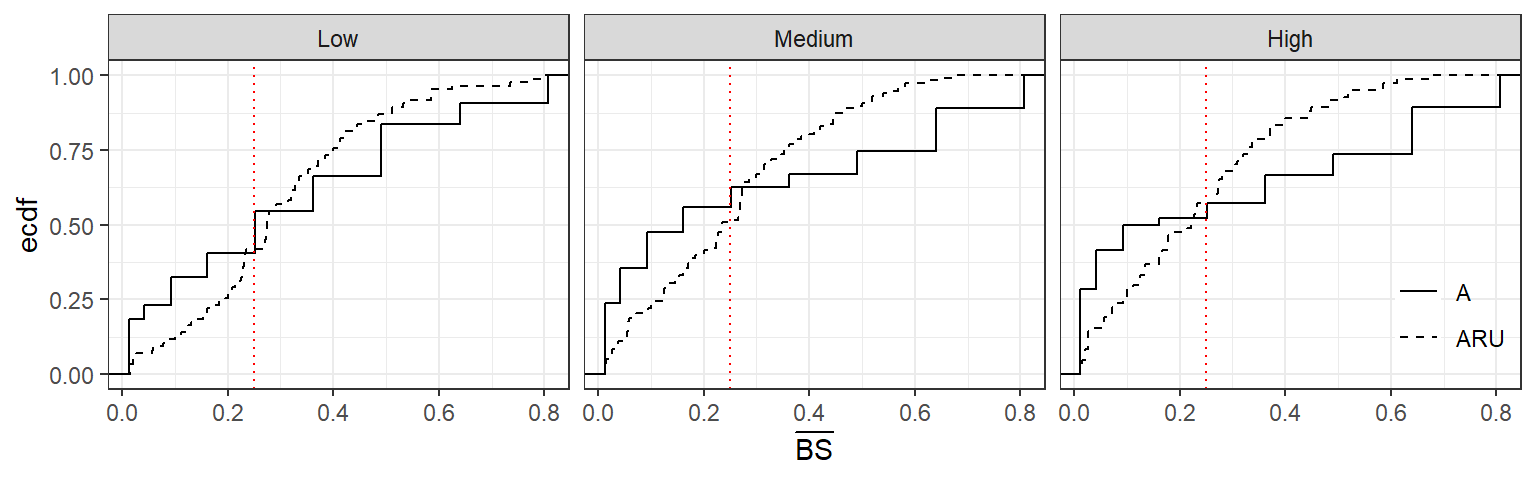}
            \caption{ARU}
            \label{fig:BS_comp_ARU}
        \end{subfigure} 
              \begin{subfigure}[b]{0.8\textwidth}
            \includegraphics[width=\textwidth]{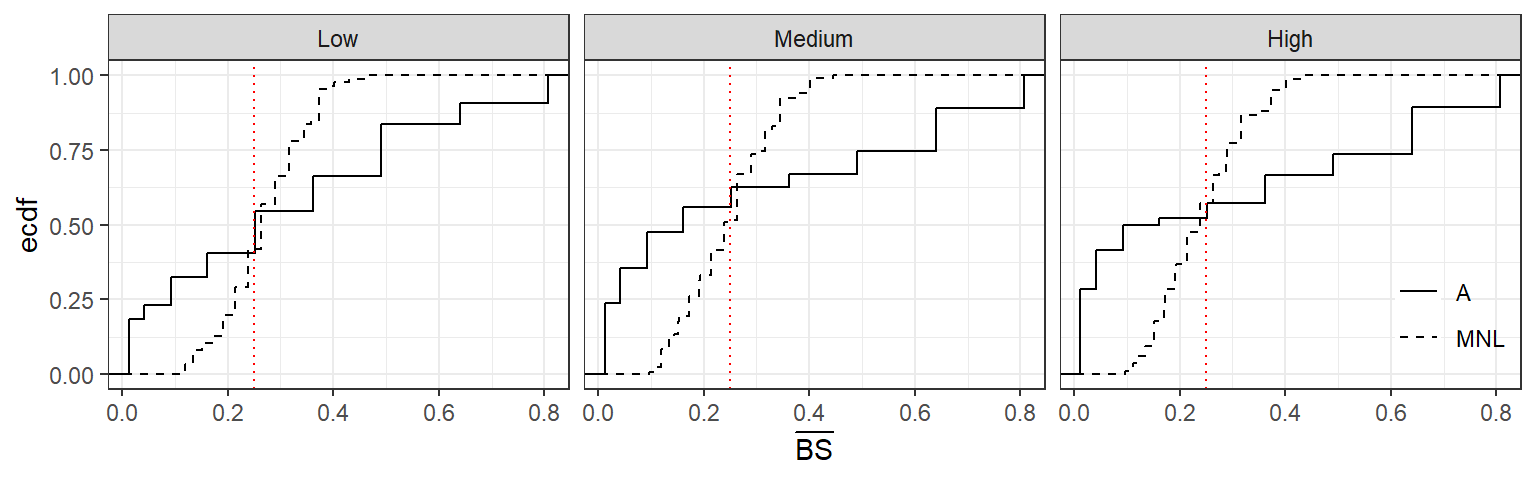}
            \caption{MNL}
            \label{fig:BS_comp_MNL}
        \end{subfigure}
            \caption{Cumulative distributions of Brier scores for direct assessment (solid line) vs.\ EUM, ARA, ARU, and MNL recompositions 
            (dotted lines) by level of knowledge (low, medium, high).}
    \label{fig:BS_by_knowledge}
\end{figure}

\begin{figure}[p]
    \centering
        \begin{subfigure}[b]{0.8\textwidth}
            \includegraphics[width=\textwidth]{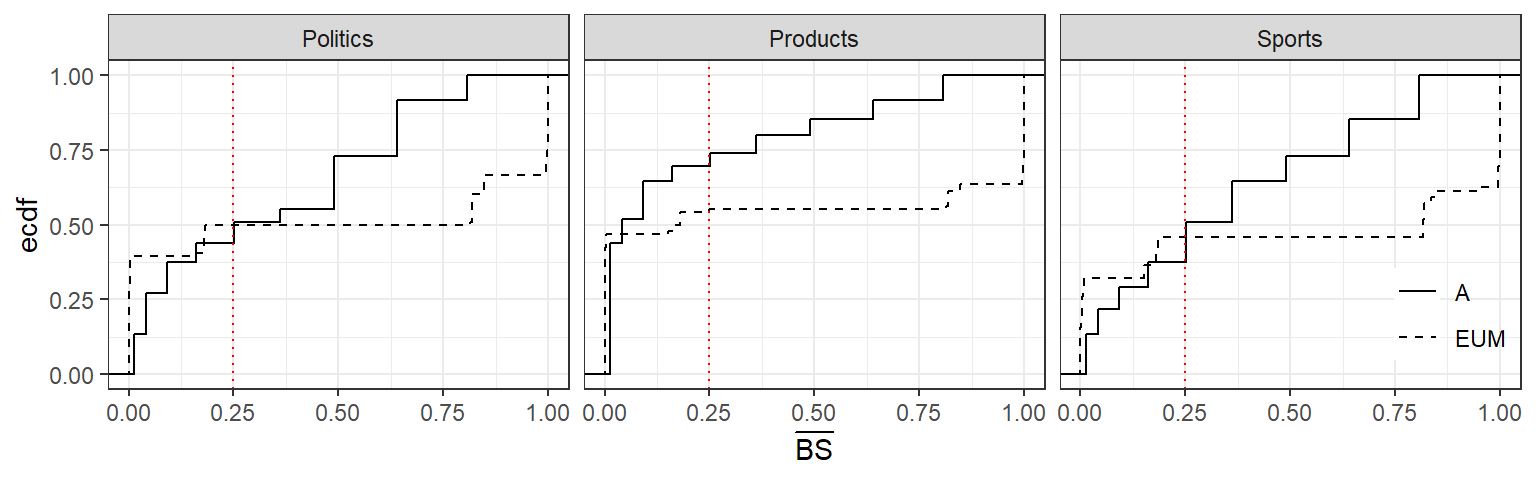}
            \caption{EUM}
            \label{fig:BS_comp_EUM_by_q}
        \end{subfigure} 
        \begin{subfigure}[b]{0.8\textwidth}
            \includegraphics[width=\textwidth]{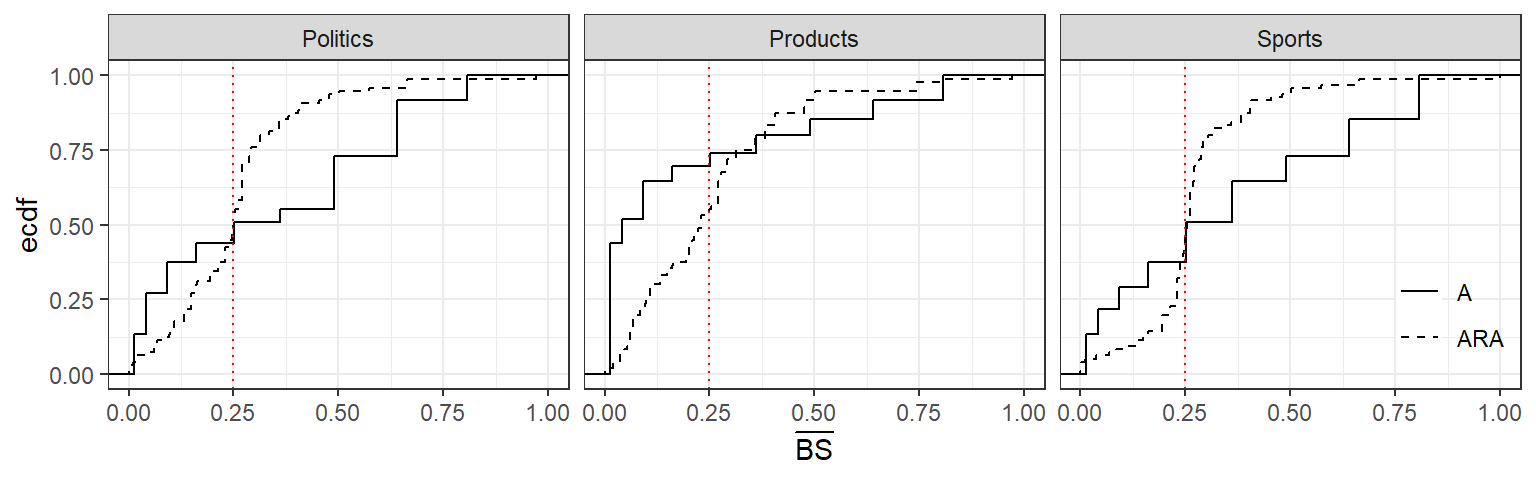}
            \caption{ARA}
            \label{fig:BS_comp_ARA_by_q}
        \end{subfigure} 
        \begin{subfigure}[b]{0.8\textwidth}
            \includegraphics[width=\textwidth]{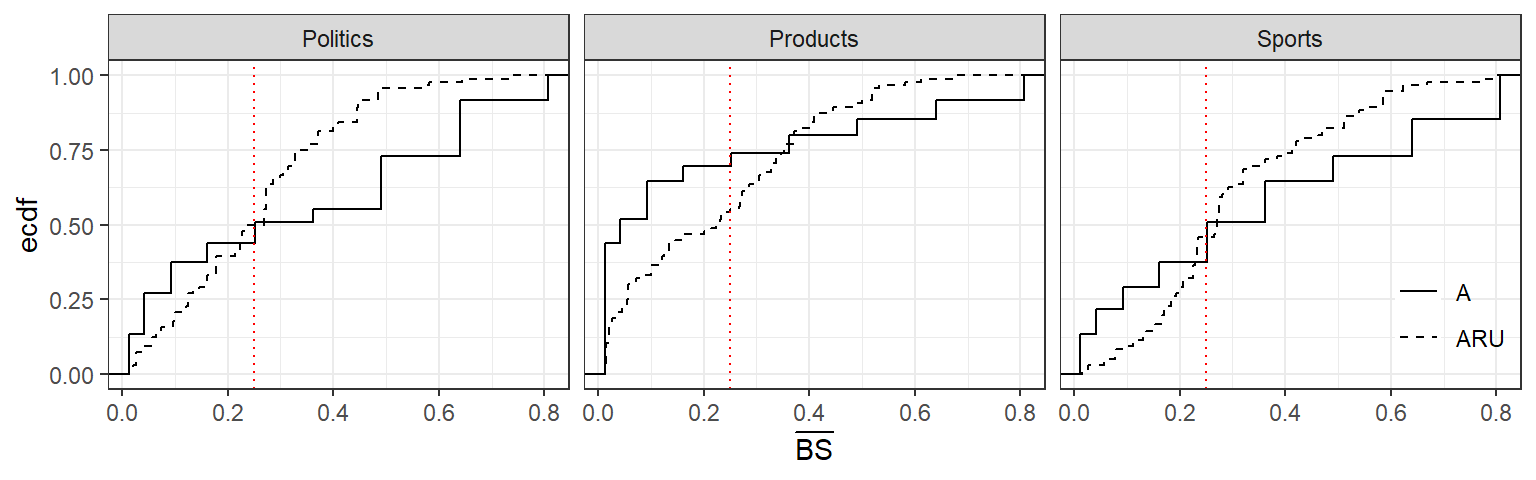}
            \caption{ARU}
            \label{fig:BS_comp_ARU_by_q}
        \end{subfigure} 
              \begin{subfigure}[b]{0.8\textwidth}
            \includegraphics[width=\textwidth]{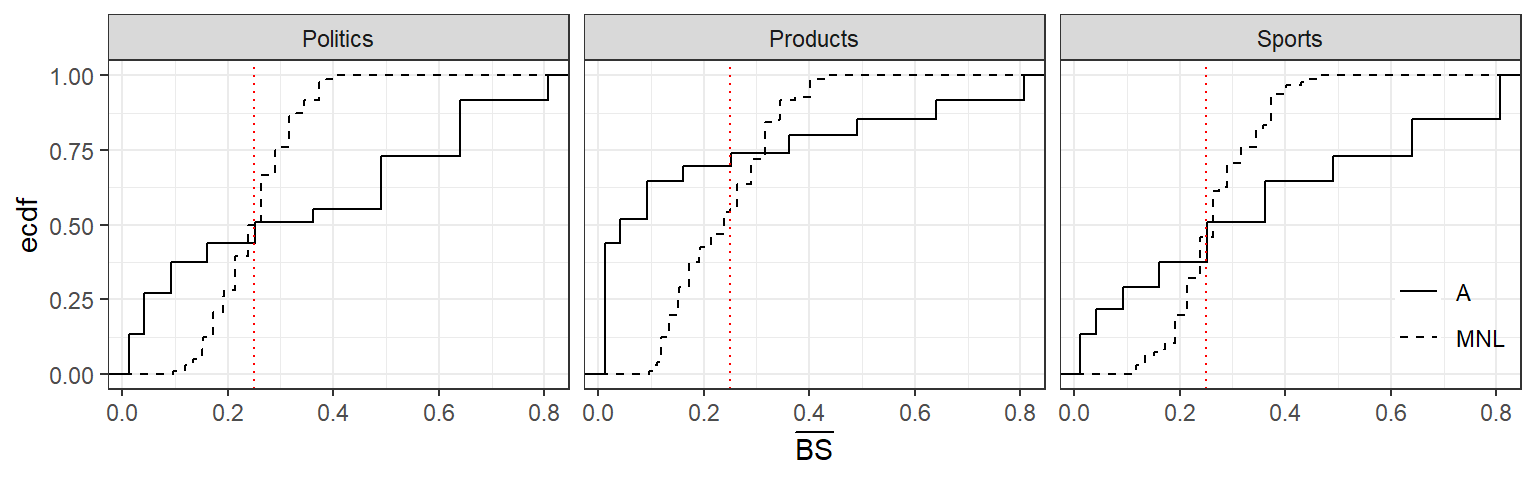}
            \caption{MNL}
            \label{fig:BS_comp_MNL_by_q}
        \end{subfigure}
            \caption{Comparison of Brier score cumulative distribution of direct assessment (solid line)  vs.\ recompositions (dotted lines) per forecasting question (politics, products, sports). }
    \label{fig:BS_by_question}
\end{figure}

We first compare the Brier score of the direct probability assessments
of $p_A$ and those of the four recompositions using their empirical cumulative distribution
functions (cdf). Since we expect the $p_A$ Brier scores to be affected by both the level of knowledge and the forecasting question, we show these comparisons separately according to these factors in 
Figures~\ref{fig:BS_by_knowledge} (knowledge) and~\ref{fig:BS_by_question} 
(question), with solid lines referring to direct assessment Brier score cdfs  and 
dotted ones to those based on recomposition. 

Figure~\ref{fig:BS_by_knowledge} shows how subjects more knowledgeable  on a forecasting question obtain better Brier scores in their direct assessment of $p_A$. The empirical cdfs  suggest that Brier scores associated with direct assessment 
have more mass concentrated around $0$ as the level of knowledge increases (towards the right of the figure). The solid lines towards the left are also closer to the diagonal indicating that Brier scores from direct assessments are more uniformly distributed between $0$ and $1$
for less knowledgeable participants. 
In contrast, Brier scores cdfs from the EUM method 
make bigger jumps near 0 and 1 leading to a bimodal behaviour, regardless of the declared knowledge level, suggesting that this method is too extreme, i.e., leading to binary prediction that will get the best Brier score if right but the worse if wrong.
Note also that the EUM cdf is essentially stochastically dominated by the direct assessment
cdf, indicating better (lower) Brier scores are more frequent with direct assessments.  
For the other three recomposition methods, the Brier scores accumulate (dotted lines) faster over lower values as the knowledge level increases, suggesting that the methods operate better with more knowledgeable subjects. Globally, the recompositions seem to improve the 
performance of worse respondents and harm that of the best respondents, with this effect
more intense for MNL than for ARU and, in turn, than for ARA.
We also appreciate, Figure~\ref{fig:BS_by_question}, how the question referring to \textit{Products} seems easier than the other two, with the corresponding lines accumulating more mass on lower Brier score values, and significantly concentrated closer to $0$. The cumulative distributions in this figure suggest again that the EUM recomposition seems too extreme and that recompositions seem to operate better with difficult questions.  

When comparing direct assessment and the four recompositions, we observe that direct assessment seems better predictively than the recomposition based on \acrshort{eum}; the two modes in relation with the recomposition, appreciated through the rapid 
growth of the cdf near 0 and 1, suggest that such model is too extreme. 
However, the situation improves for the ARA, ARU and MNL recompositions which 
display better predictive capabilities than direct assessment. For example, 
panel (b) of Figure~\ref{fig:BS_by_knowledge} shows how in a situation in which 
a forecaster self-identifies as non-knowledgeable, the ARA recomposition corrects the forecast towards $0.5$ which in turns center the Brier scores around $0.25$. 
This reduces the dispersion of forecasts from less knowledgeable subjects 
leading to a lower average Brier score.
Similarly, for subjects identified to have more knowledge the ARA recomposition is able to correct their forecasts towards more accurate ones, as appreciated on the right of the panel 
by the dotted line stochastically dominating the solid one for higher (worse) Brier scores (i.e., a cdf is better when is on the left of the other cdf). 

\begin{figure}[htbp!]
    \centering
    \includegraphics[width=1\textwidth]{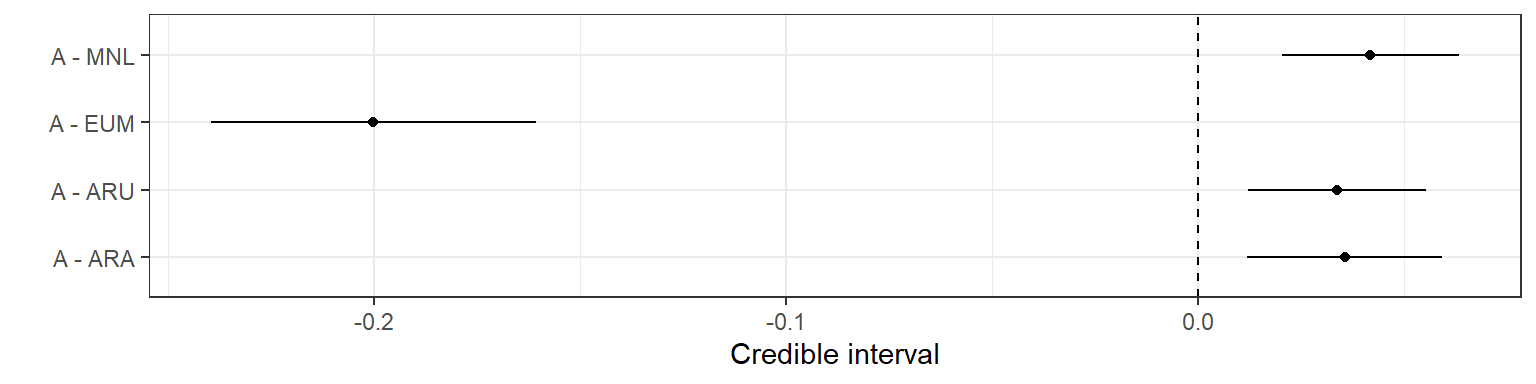}
    \caption{Credible intervals for differences
    in Brier scores between direct assessment and the four recomposition methods.}
    \label{fig:bs_tests}       
\end{figure}

These explorations are confirmed through hypothesis tests 
aimed at checking whether Brier scores attained via direct assessment are different than those attained 
via the recompositions under study. 
Assuming non-informative priors, Figure \ref{fig:bs_tests} presents 95\% 
credible intervals  for the difference 
in Brier score obtained with direct assessments $p_A$ and recomposed assessments $\hat{p}_A$. In the case of the \acrshort{eum} recomposition, 0 is to the right of the interval and, therefore, worse predictive capabilities are attained, suggesting 
that this rule is very extreme and does not improve direct forecasts, in spite 
of enforcing consistency. On the other hand, for the ARA, ARU and AML  recomposition methods, 0 is to the left of the intervals and, therefore, we obtain better, more accurate, predictions through these recompositions than using direct assessment. This suggests that recompositions
based on ARA, ARU and MNL models, constitute a relevant way for analysing adversarial belief formation and designing effective \acrshort{sej} methods in strategic situations, as they both ensure 
coherence and seem to improve predictive capabilities.

\begin{figure}[htbp!]
    \centering
    \includegraphics[width=0.8\textwidth]{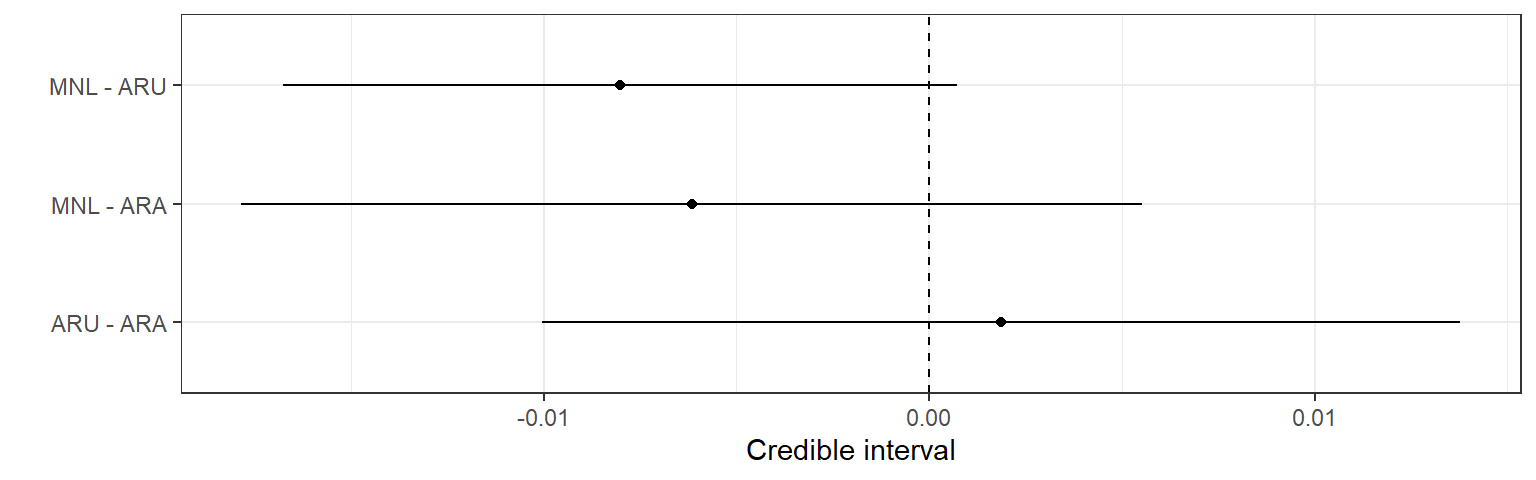}
    \caption{Credible intervals for Brier scores differences between 
    \acrshort{mnl}, \acrshort{aru} and ARA recompositions.}
    \label{fig:bs_tests2}       
\end{figure}

It is then natural to compare the performance among the three better methods (ARU, ARA 
and MNL) to find out if any of them stands out over the others. Figure \ref{fig:bs_tests2} displays 95\% credible interval for the comparison between every pair of these three recompositions using non-informative priors. The intervals suggest no major 
advantage, except perhaps a minor one of MNL over ARU.

Note that, referring to Figures~\ref{fig:BS_by_knowledge} and ~\ref{fig:BS_by_question},
the proportion of forecasting questions with a Brier score bigger than 0.25 (the one attained with random guessing) is quite important (around 50\% for all four recomposition methods). This means that although we improved upon direct assessment, the predictive quality is not that good. However, this may be due to the initially 
poor predictive capabilities of the involved subjects. 
To illustrate a possible amelioration approach, we consider a coherence based procedure which selects only those experts which were 
coherent in all three questions and then aggregate the recomposed estimates of the selected participants by averaging. Such synthetic coherent participant obtained a Brier score of $0.194$. This type of approach of selecting better experts is one of the general principles in the Good Judgement approach in~\cite{mellers2015identifying}.


\subsection{Assessing the impact of reflection}

As a final stage, we study the effect of the reflection induced by our decomposition tasks in the respondents' direct forecasts. In our experiment, 69\% of the participants did not update their direct assessments after reflection. Thus, the question here is whether there is improvement in the predictive capability for respondents who actually changed their direct probability judgments after reflection. 

\begin{figure}[htbp!]
    \centering
    \includegraphics[width=0.5\textwidth]{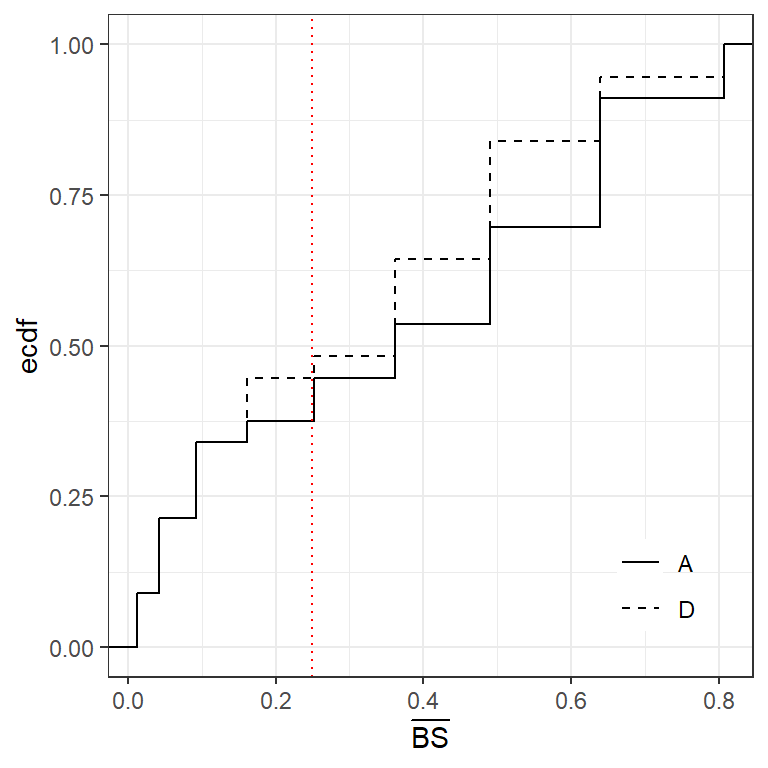}
    \caption{Empirical distributions of the Brier scores for prediction probabilities $p_A$ and $p_D$.}
    \label{fig:bsad}
\end{figure}

\begin{figure}[htbp!]
    \centering
    \includegraphics[width=0.6\textwidth]{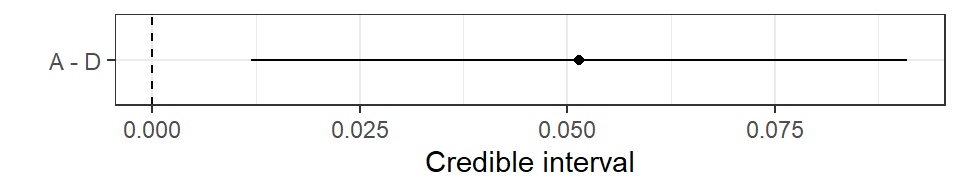}
    \caption{Credible interval on Brier score differences between $p_A$ and $p_D$.}
    \label{fig:bsd_test}       
\end{figure}

To answer this, Figure~\ref{fig:bsad} provides the Brier score empirical cdfs of the predictions based on the direct assessments before and after going through Tasks $2$ and $3$, for the cases in which $p_A$ and $p_D$ estimates were different. The Brier scores for the $p_D$ predictions obtained after reflection seem to stochastically dominate those corresponding to their initial $p_A$ predictions, indicating that better (lower) Brier scores are more frequent after reflection. Figure~\ref{fig:bsd_test} confirms this by providing the 95\% credible interval for the average difference in Brier scores between $p_A$ and $p_D$, with a non informative prior. We note that $0$ is clearly not in such interval, with an average difference of $0.05$ in the Brier score, supporting the hypothesis that the reflection induced by the proposed decomposition tasks had a relevant effect in improving forecasting accuracy after reflection when direct elicitation is revised.

\section{Discussion}
\label{sec:cap3discussions}

Forecasting adversarial actions is frequently needed in
important domains including cybersecurity,
 national security, intelligence, and competitive business, to name but a  few. This is a complex problem as 
we need to take into account the strategic features of the 
adversaries. However, we do not have these directly available to try to assess their preferences, beliefs and concept solution approaches to  predict their actions. 

Therefore, we have introduced an approach to facilitate such strategic forecasts
creating a model of the adversary's decision problem to elicit a probability distribution of their 
  decisions. Assuming that the adversary is an expected utility maximizer (or has some other criterion, as in prospect theory), we try to assess his beliefs and preferences. If these were known, we could identify his optimal action. However, our uncertainty about the adversaries' beliefs and preferences is propagated to his decisions through our adversary's decision model, leading to a probability distribution over his actions. This approach can thus be framed as a tool for \acrshort{sej} elicitation when we need to deal with probabilities referring to actions by opponents.
In it, our uncertainty on the adversary decisions is a consequence of our uncertainty on adversary's beliefs and preferences involved in the his decision problem. However, we still need to take into account the uncertainty that we might have about the adversary's rationality, designated concept uncertainty, beyond the traditional epistemic and aleatory uncertainties in risk analysis \citep{Banksy}. For this we have developed four behavioural 
recompositions incorporating the basic ingredients obtained through ARA decompositions.
 These are interpretable as a result of cognitive biases, decision making heuristics or errors in the implementation of the decision making process. 
 Three of the recompositions (ARU, MNL and ARA) have shown effective for adversarial forecasting purposes, enforcing coherence, improving upon direct assessment as well as improving when properly aggregated among selected experts,
 as empirically validated with data from a face-to-face behavioural experiment. Specifically, the results of this experiment suggest the following conclusions:
\begin{enumerate}
    \item We have identified in our analysis respondents with clear inconsistencies between the probabilities assessed directly and their revised ones after going through a decomposition exercise. For those who revised and corrected their assessments, we observed some improvement in their forecasting performance. This demonstrates the value of reflection through decomposition and the importance of updating judgments.
    \item We have generated recomposed probabilities from the decomposed assessments through four different recomposition models for the adversary's choice:  EUM, ARA, ARU and MNL. By construction, the proposed recompositions allow us to enforce coherence.
    \item However, not all them seem equally relevant in predictive terms, as per their Brier scores. In particular,  better results are attained in terms of improving predictive capabilities using 
    the \acrshort{aru}, \acrshort{mnl} and {\acrshort{ara}}  recompositions than {\acrshort{eum}} recomposition.
    \item A careful selection of participants leads as well to improved predictive capabilities
    in Brier score terms.
\end{enumerate}


These results highlight the large cognitive burden and complexity  required when forecasting adversarial actions, 
which thus seems a task 
pertaining to the conscious and analytic System 2, in
\cite{kahneman2012think}'s dual thinking terminology. As 
the fast and automatic System 1 frequently overcomes System 2, direct adversarial belief elicitation does not take into account strategic aspects, providing worse  estimations.
Therefore, a promising strategy that may improve adversarial belief formation is to perform externally those tasks that would have corresponded to the System 2, but it has not actually performed. This is the approach proposed 
with our decomposition-recomposition methods which focused on (i)  accurate elicitation of the beliefs to decompose 
the basic building blocks of the adversary's decision analysis; and (ii) 
recomposition of the beliefs by 
 integrating a behavioural perspective.  

The results in this paper contribute to our understanding of  belief formation in adversarial situations. We have shown that an agent’s probability distribution on the actions to be chosen by an adversary, as obtained by direct elicitation, differs from that obtained 
through behavioural recomposition of ARA 
decompositions. Such difference remains even when 
 direct elicitation is performed after the reflection exercise 
 entailed by eliciting her beliefs on each individual elements of her adversary's decision problem.

Relevant open research issues include the consideration of other behavioural models to be integrated in the \acrshort{ara} model and capable of increasing the accuracy of the estimates: cognitive biases related to uncertainty processing (such as probability insensitivity or formation of decision weights) or decision heuristics (such as anchoring). 
Further research is required to develop effective behavioural recomposition methods for specific fields, such as in cybersecurity risk analysis, and validate them with larger size field and/or behavioural experiments. There is also the need for further experiments comparing the accuracy of our approach with alternative methods to forecast adversaries' decisions. These alternative forecasting methods may include active role-playing, for which there is empirical evidence of good performance when compared against game-theoretic and unaided judgmental predictions \citep{green2002roleplaying, green2005roleplaying}, as well as other forecasting techniques identified by \cite{singer1990forecasting} for predicting business competitors' actions.


\appendix

\section{Screenshots from experiment}
\label{sec:cap3screenshots}

This appendix shows the screenshots of the experiment for the first of the three decisions considered in the experiment (Theresa May's decision of calling for elections). The screenshots are in Spanish, since the experiment was run in Spain (no version of the experiment in English was developed).

\begin{figure}[H]
\centering
\includegraphics[width=0.9\textwidth]{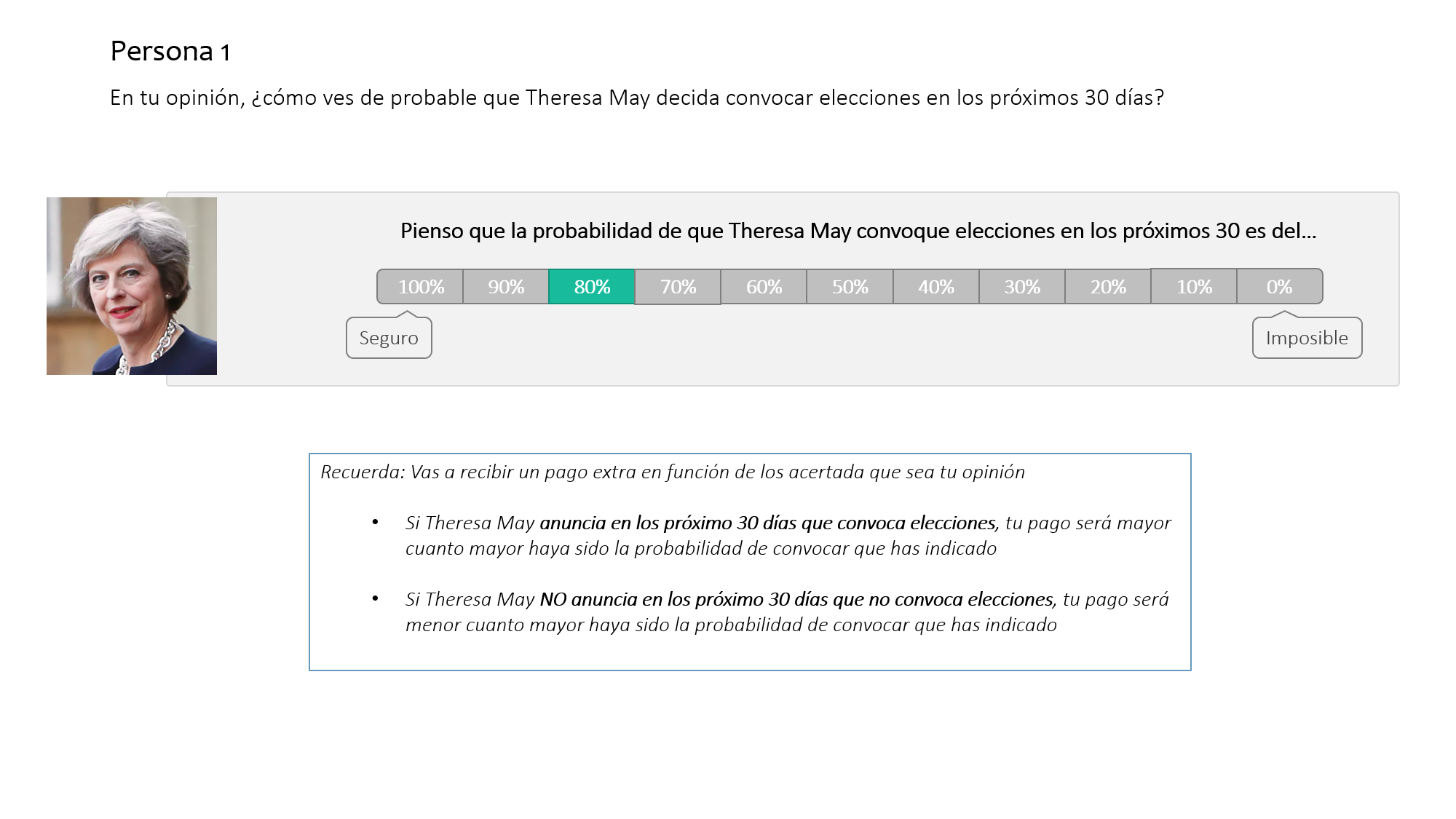}
\caption{Disclosure of $p_A$.}
\label{fig:taska}       
\end{figure}

Figure \ref{fig:taska} presents the screen used to directly assess the probability ($p_A$ and $p_D$) of Theresa May calling for elections in 30 days, as a 10\% width probability interval.

\begin{figure}[H]
\centering
\includegraphics[width=0.9\textwidth]{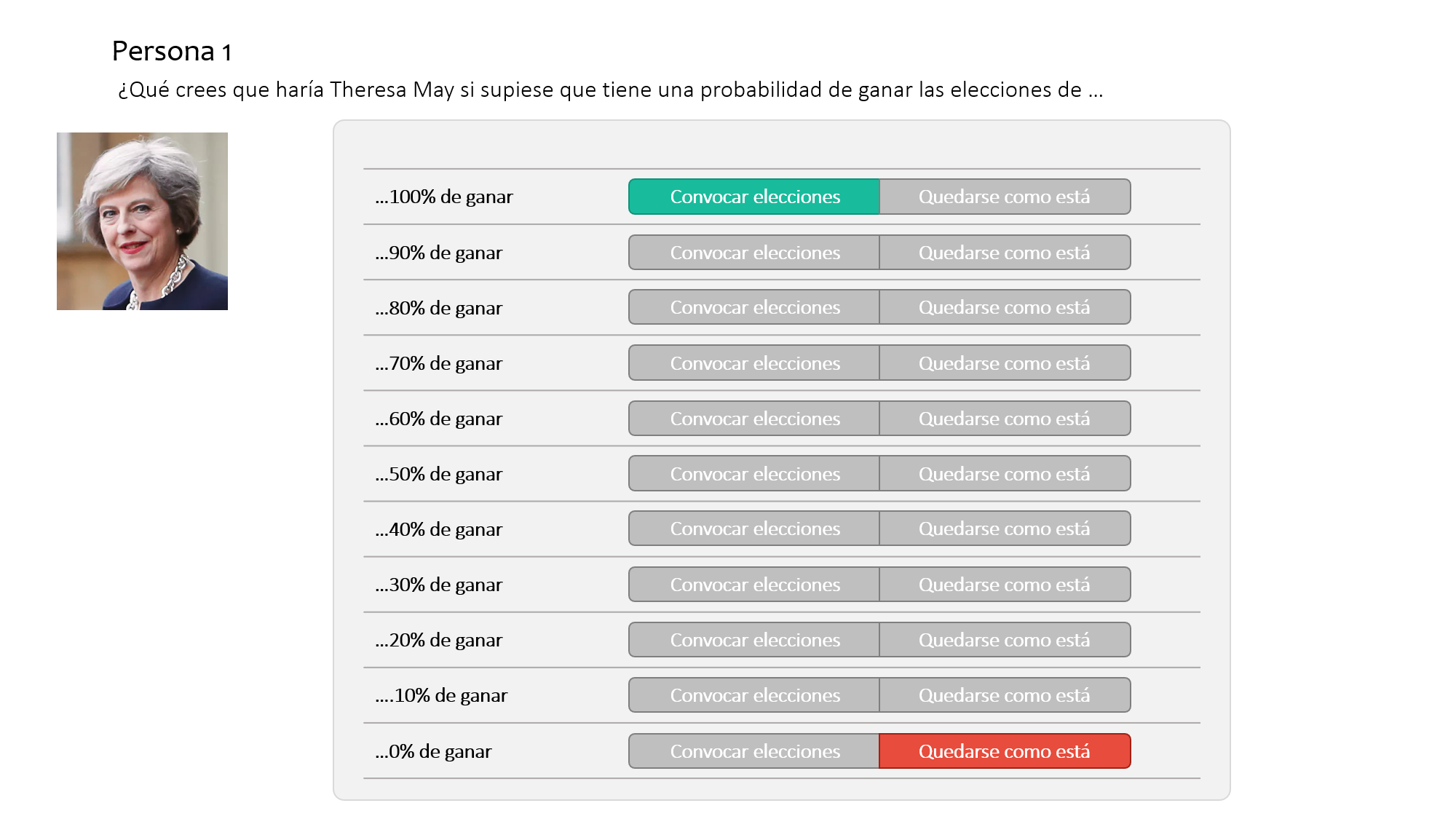}
\caption{Disclosure of $p_B$.}
\label{fig:taskb}       
\end{figure}

Figure ~\ref{fig:taskb} presents the screenshot for the second task, where again, we identify a 10\% width interval for the required probability.

\begin{figure}[H]
\centering
\includegraphics[width=0.9\textwidth]{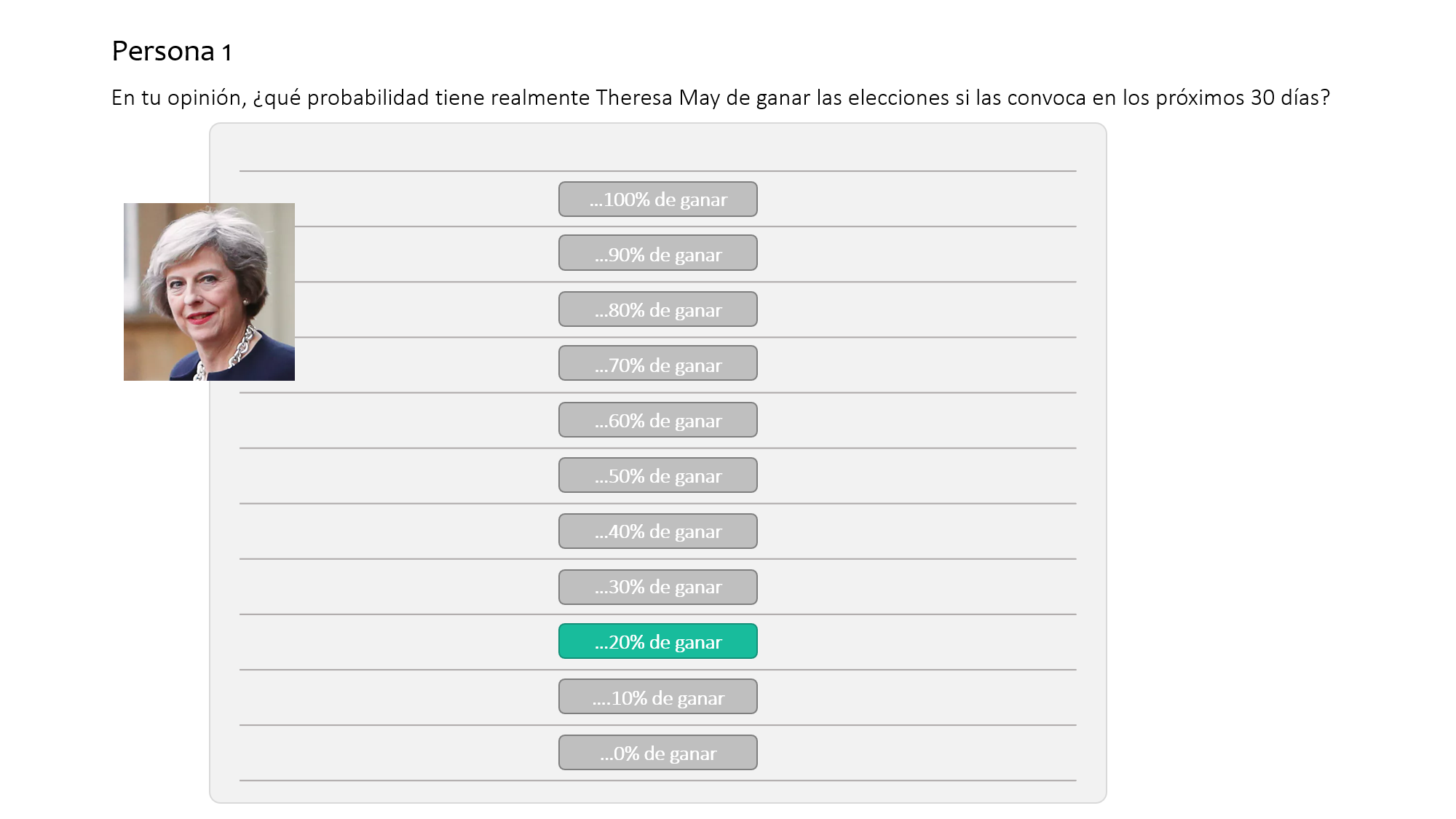}
\caption{Disclosure of $p_C$.}
\label{fig:taskc}       
\end{figure}

Figure \ref{fig:taskc} presents the screenshot for the third task.

\section{Handling imprecision in experiment responses }
To acknowledge the imprecision in the responses present in the 
proposed tasks, expressed through 10\% probability intervals,
we fit beta distributions  
    with quantiles in the extremes of the corresponding 
    interval having a probability of 0.9 and median in the midpoint of the interval.  
    As an example, participant 8 selected the option 30\% in Task 1 for the first question; we assimilate it with the
    interval [0.25, 0.35] having probability 0.9 and median 
    0.3 which,
    using routine \texttt{get.beta.par} from R, leads to
    modelling the uncertainty over such 
    probability through a $\beta eta (68.08, 158.56)$ 
    distribution. Such distributions served  to sample observations whenever requested
    in the analysis as described in the main part of the paper.

\section{Distribution of features for respondents}
\label{sec:cap3sample}

The respondents' profile distribution is shown in Table ~\ref{tab:1}. Three quarters of them were public/private workers and were almost equally distributed between both genders. Regarding education level of participants, most of them had either some years of university or a university degree.

\begin{table}[htbp!]
\centering
\begin{tabular}{lrr}
\textbf{Profile}                         & \textbf{N}  & \textbf{\%}    \\ \hline
\textit{Total}                           & \textit{96} & \textit{100.0} \\ \hline
18-35                                    & 26          & 27.1           \\
36-50                                    & 39          & 40.6           \\
51-74                                    & 31          & 32.3           \\ \hline
Male                                     & 50          & 52.1           \\
Female                                   & 46          & 47.9           \\ \hline
Primary education                        & 19          & 19.8           \\
Secondary education                      & 28          & 29.2           \\
Tertiary education                       & 49          & 51.0           \\ \hline
Freelance                                & 8           & 8.3            \\
Public/private worker                    & 72          & 75.0           \\
Unemployed                               & 7           & 7.3            \\
Other                                    & 9           & 9.4            \\ \hline
\end{tabular}
\caption{Distribution of participants by age, gender, education and employment.}
\label{tab:1}
\end{table}



\subsubsection*{Acknowledgments}

Research supported by the EU's Horizon 2020 project 740920 CYBECO (Supporting Cyberinsurance from a Behavioural Choice Perspective). DRI is grateful to PID2021-124662OB-I00 from the Ministry of Science and Technology project,  
  the AXA-ICMAT Chair in Adversarial Risk Analysis
  and 
  EOARD (FA8655-21-1-7042) and 
  AFOSR (FA-9550-21- 1-0239)  projects. This work was partially supported by the National Science Foundation under Grant DMS-1638521 to the Statistical and Applied Mathematical Sciences Institute and a BBVA Foundation project (AMALFI). 
  Jose Vila is grateful to the financial support by Conselleria de Innovación, Universidades, Ciencia y Sociedad Digital of the Generalitat Valenciana under the Excellence Program PROMETEO 2023 - CIPROM/2022/029 and project PID2021-127946OB-I00 funded by MCIN/AEI/10.13039/501100011033 and UE-FEDER.



\bibliographystyle{abbrvnat} 

\bibliography{references}

\end{document}